\documentclass[aps,prb,showpacs,superscriptaddress,amsmath,twocolumn]{revtex4-1}

\usepackage{amsmath}
\usepackage{amssymb}
\usepackage{graphicx}
\usepackage[caption=false]{subfig}

\begin{document}

\title{Frequency dependent magneto-optical conductivity in the generalized $\alpha - T_3$ model}

\author{\'Aron D\'aniel Kov\'acs}
\affiliation{Department of Physics of Complex Systems, E\"otv\"os University, H-1117 Budapest, P\'azm\'any P{\'e}ter s{\'e}t\'any 1/A, Hungary}

\author{Gyula D\'avid}
\affiliation{Department of Atomic Physics, E{\"o}tv{\"o}s University, H-1117 Budapest, P\'azm\'any P{\'e}ter s{\'e}t\'any 1/A, Hungary}

\author{Bal\'azs D\'ora}
\affiliation{Department of Theoretical Physics and BME-MTA Exotic  Quantum  Phases Research Group, Budapest University of Technology and
  Economics, Budapest, Hungary}

\author{J{\'o}zsef Cserti} 
\affiliation{Department of Physics of Complex Systems, E\"otv\"os University, H-1117 Budapest, P\'azm\'any P{\'e}ter s{\'e}t\'any 1/A, Hungary}


\begin{abstract}

We have studied a generalized three band crossing model in 2D, the generalized $\alpha - T_3$ lattice, ranging from the pseudospin-1 Dirac equation through a quadratic+flat band touching to the
pseudospin-1/2 Dirac equation. A general method is presented to determine the operator form of the Green's function, being gauge and representation independent.
This yields the  Landau level structure in a quantizing magnetic field and the longitudinal and transversal  magneto-optical conductivities
of the underlying system
Although the magneto-optical selection rules allow for many transitions between Landau levels, the dominant one stems from exciting a particle from/to the flat band to/from a propagating band.
The Hall conductivity from each valley is rational (not quantized at all), in agreement with Berry phase considerations, though their sum is always integer quantized.

\end{abstract}


\maketitle

\section{Introduction}
\label{intro:sec}

Since the first isolation of graphene\cite{castro07} in  2004 and the theoretical prediction and experimental realization of topological insulators\cite{hasankane,zhangrmp}, the Dirac equation and its variants
have started to attract almost unprecedented attention in condensed matter and related fields. The peculiar spinor structure of the Dirac equation, which e.g. stems from the two sublattices of the 2D honeycomb
lattice in graphene, gives rise to many topology related phenomena such as a Berry phase\cite{castro07} of $\pi$, unusual Landau quantization in a magnetic field and the related unconventional quantum Hall effect\cite{novoselov2},
just to mention a few immediate consequences.

The 2D massless Dirac equation possesses the deceivingly simple form as 
\begin{gather}
H_{S=1/2}=v_F {\bf{Sp}}=v_F\left[\begin{array}{cc}
0 & p_-\\
p_+& 0
\end{array}\right],
\label{dirac12}
\end{gather}
 where $v_F$ is the Fermi velocity of the underlying system and plays the role of the effective speed of light, ${\bf p}=(p_x,p_y)$ is the 2D momentum,
 $p_\pm=p_x\pm i p_y$
 and $\bf S$ stands for the spin-1/2 Pauli matrices, which represent the sublattice degree of freedom in this instance.
Shortly after the discovery of graphene, this equation was generalized, still in 2D, to arbitrary pseudospin-$S$, 
known as the Dirac-Weyl equation  with $\bf S$ now representing the $(2S+1)\times (2S+1)$ matrix representations of the $SU(2)$ algebra,
and several lattices have been proposed, hosting these Weyl fermions\cite{greens1,urban,PhysRevA.80.063603,lan,watanabe}.

Similarly to other spin-$S$ problems, cases with integer and half-integer spin differ from each other.
The ensuing  spectrum consists of coaxial Dirac cones, crossing each other at the same Dirac point, and for integer
spins, an additional dispersionless flat band also shows up and crosses the Dirac point.

\begin{figure} 
\includegraphics[scale=0.6]{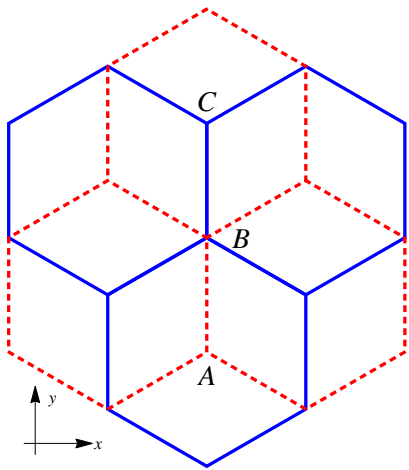}
\caption{
\label{geo_layers:fig} 
(Color online) 
The dice lattice with $t_1$  and $t_2$ hopping amplitude along the red dashed and blue solid lines. 
The on-site energy of the sixfold connected site $B$ is $\epsilon_0$.
There are three atoms $A, B$ and $C$ in each unit cell. 
} 
\end{figure}

The simplest integer spin case is the pseudospin-1 Weyl equation.
It has a $3\times 3$ matrix structure as
\begin{gather}
H_{S=1}=v_F\left[\begin{array}{ccc}
0 & p_- & 0\\
p_+& 0 & p_-\\
0 & p_+& 0
\end{array}\right],
\label{dirac1}
\end{gather}
and
in comparison to Eq. \eqref{dirac12}, many more new terms can be added to this and masses can be opened in several distinct ways\cite{PhysRevB.84.195422}.
As detailed below, Eq. \eqref{dirac1}  can be
realized in the dice or $T_3$ lattice,
composed of two 2D honeycomb lattices, which share one sublattice and is sketched in Fig. \ref{geo_layers:fig}.
Experimentally, the dice lattice can be realized 
from a trilayer structure of the face-centred cubic lattice, grown in the [111] direction~\cite{PhysRevB.84.195422}.

Recently, a novel variant of the $T_3$ lattice structure was proposed, coined as the $\alpha-T_3$ model, suggested first by Raoux \textit{et al.}~\cite{PhysRevLett.112.026402}.
Due to the three non-equivalent lattice sites of the $T_3$ lattice, two nearest neighbour hopping integrals are possible, which, however, need 
not be equal to each other.
The generalized $\alpha - T_3$ model is described alternatively by a lattice consisting of three layers of triangular lattices 
with basis atoms $A, B$ and $C$ and with only intersublattice hoppings between adjacent layers shown in Fig.~\ref{geo_layers:fig}. 
The three band tight-binding Hamiltonian in the basis  $A, B$ and $C$  is given by~\cite{PhysRevLett.81.5888,PhysRevA.80.063603,PhysRevB.84.195422,PhysRevB.90.045310}
\begin{align}
\label{Ham_TB:eq}
H_{\text{dice}} &=\left(
\begin{array}{ccc}
0 & t_1 \, f({\boldsymbol{k}}) & 0 \\ 
t_1 \, f^{*}({\boldsymbol{k}}) &\epsilon_0& t_2 \, f({\boldsymbol{k}}) \\
0 & t_2 \, f^{*}({\boldsymbol{k}}) & 0  \\
\end{array}
 \right),
\end{align}
where $t_1$ and $t_2$ are the hopping amplitudes between adjacent triangular lattice,  and we have also generalized it further by adding\cite{PhysRevB.90.045310} 
an on-site energy term $\epsilon_0$ in the middle layer, arising from, e.g., 
a real chemical potential, while $f({\boldsymbol{k}}) = 1+ 2 \exp(i 3 k_y a/2) \cos(\sqrt{3} k_x a/2)$
with ${\boldsymbol{k}} = (k_x,k_y)$ and $a$ is the nearest neighbor distance in the dice lattice 
(the distance between sites A and B), and $*$ denotes the complex conjugation.

\begin{figure}[h]
  \centering
  \includegraphics[width=4cm]{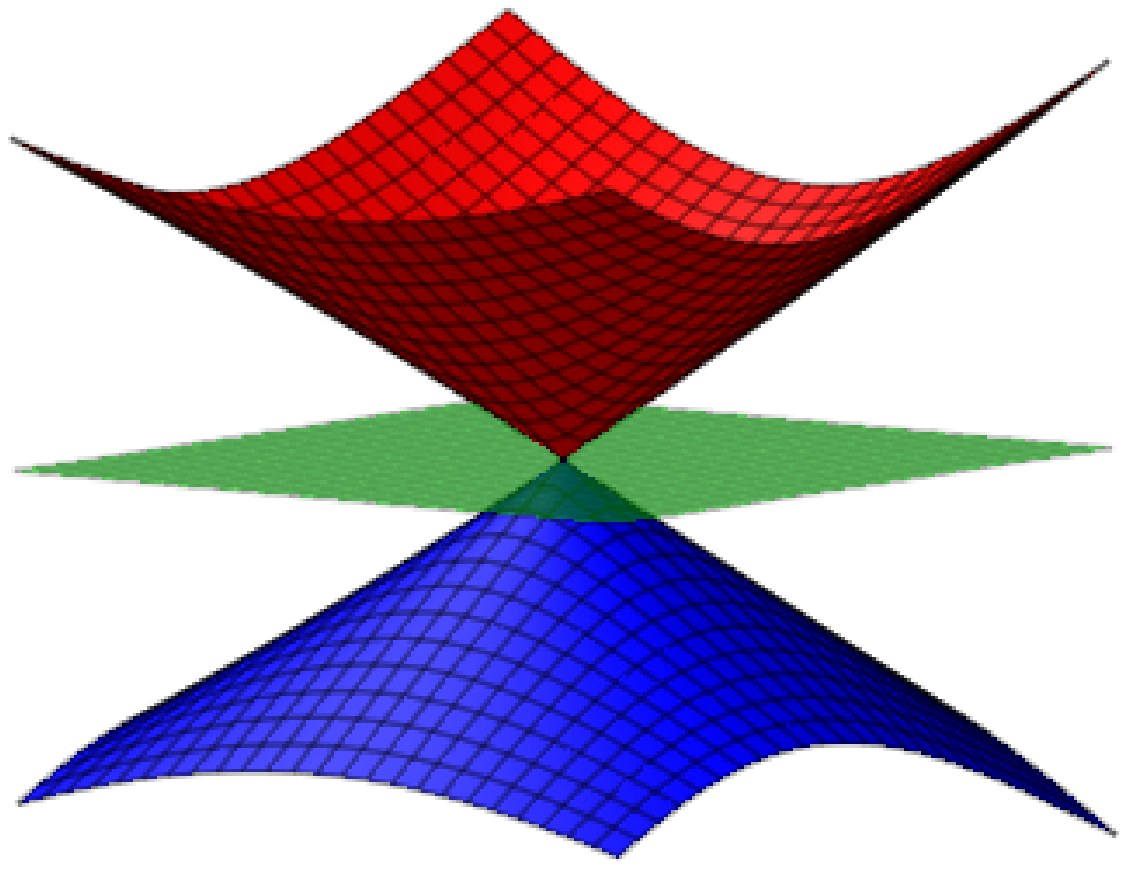}
\includegraphics[width=4cm]{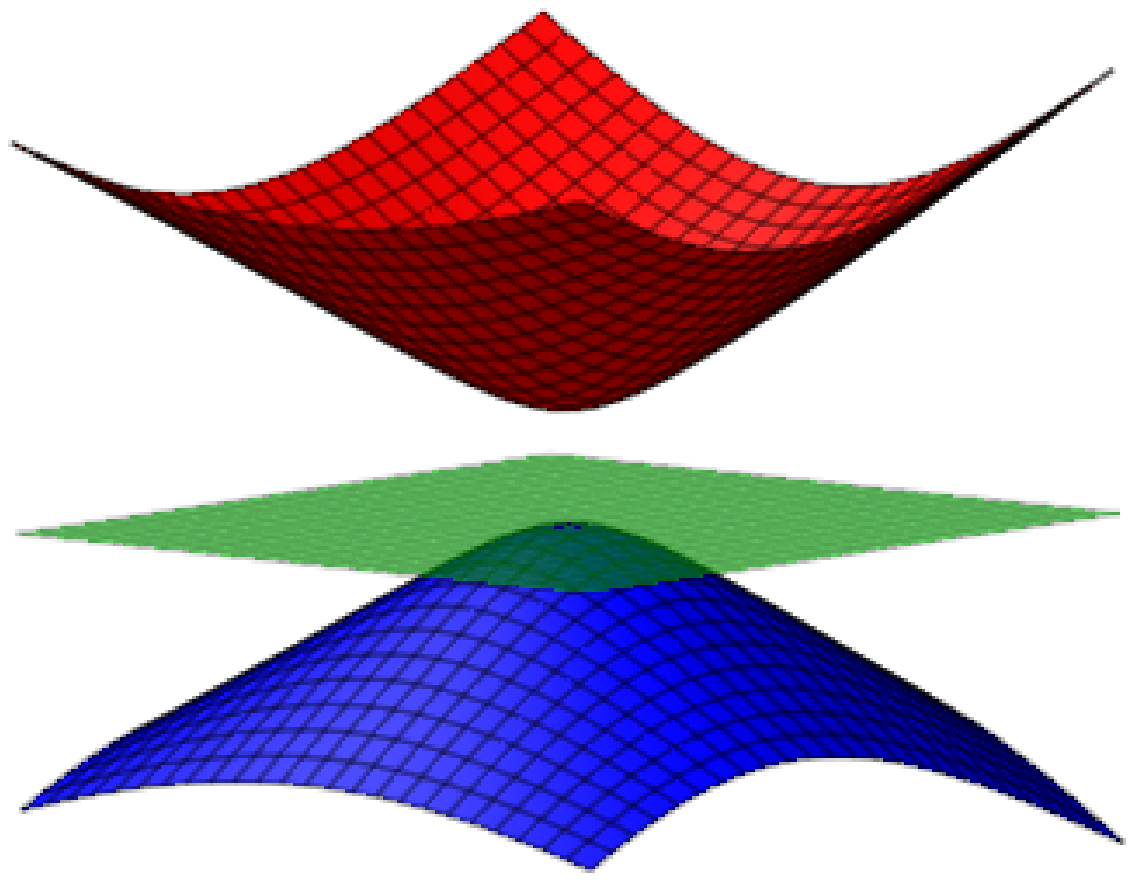}

\caption{(Color online) The energy dispersion is sketched in the low energy limit of the generalized $\alpha-T_3$ model, Eq. \eqref{eigenenergies}
for $\epsilon_0=0$ (left panel) and
$\epsilon_0>0$.}
\label{fig:disp}
\end{figure}

Linearizing the function $f({\boldsymbol{k}})$ around the $\mathbf{K}=(2\pi/3\sqrt{3}a,2\pi/3 a)$ point 
in the Brillouin zone we have
$f({\boldsymbol{K}} + {\boldsymbol{k}}) \approx (3 a/2) (k_x - i k_y)$. 
Then, the linearized form of the Hamiltonian (\ref{Ham_TB:eq}) for low energy states (around the $\mathbf{K}$ point) 
reads
\begin{equation}
\label{Kham:eq}
H_K=\frac{3a}{2}\, \left(
\begin{array}{ccc}
0 & t_1 \, k_{-} & 0 \\ 
t_1 \, k_{+} & \epsilon_0 & t_2 \, k_{-} \\
0 & t_2 \, k_{+} & 0\\
\end{array}
 \right),
\end{equation}
where $k_{\pm}=k_x\pm ik_y$. The eigenenergies are
\begin{gather}
E_0({\bf k})=0,\hspace*{4mm}
E_\pm({\bf k})=\frac{\epsilon_0}{2}\pm\sqrt{\frac{\epsilon_0^2}{4}+v_F^2 k^2},
\label{eigenenergies}
\end{gather}
where $v_F=3a\sqrt{t_1^2+t_2^2}/2$. The resulting dispersion relation is plotted in Fig. \ref{fig:disp}.

Similarly, around the $\mathbf{K}^\prime=-\mathbf{K}$ point we have 
$f({\boldsymbol{K}^\prime} + {\boldsymbol{k}}) \approx -(3 a/2) (k_x + i k_y)$ and  
thus, the Hamiltonian (\ref{Ham_TB:eq}) for $K'$ valley can be obtained by a unitary transformation with
matrix $U$ and a replacement of the parameters as  
\begin{align}
\label{uniter-U:eq}
U &=\left(\begin{array}{ccc}
0&0&1\\
0&1&0\\ 
1&0&0
\end{array}
\right), \,\,\, \mathrm{and } \,\,\, (t_1,t_2)\rightarrow (-t_2,-t_1). 
\end{align} 
Note that in case of $t_1=t_2$ and $\epsilon_0=0$ the two valleys are equivalent, 
however if either of these conditions are not met, this symmetry is broken.

When the on-site energy $\epsilon_0=0$ (Fig. \ref{fig:disp}),  there are two special cases for this generalized model:
i) for $t_1=t_2$, this equation reduces to the pseudospin-1 Dirac-Weyl model of Eq. \eqref{dirac1} (see Refs.~\onlinecite{PhysRevLett.112.026402,PhysRevB.92.245410})
and ii) for $t_2 =0 $ and $t_1 \ne 0$ (or the other way round) then it corresponds to the pseudospin-1/2 Dirac equation of Eq. \eqref{dirac12} (i.e. the graphene) and contains a completely detached flat band.

For $\epsilon_0 \neq 0$, on the other hand, the model contains two parabolic bands, separated by a bandgap of size $|\epsilon_0|$, and an additional flat band appear, 
touching the bottom or the top of one of the parabolic bands\cite{PhysRevB.90.045310},
depending on the sign of the local on-site energy  term, as follows from Eq. \eqref{eigenenergies}. See Fig. \ref{fig:disp}!


The pseudospin-1 Dirac-Weyl equation also describes the low energy excitations in a Lieb lattice, and has been realized using photonic waveguides\cite{mukherjee,vicencio,guzman}.
Recently, the DC Hall response and the optical conductivity without magnetic field of the $\alpha-T_3$ lattice were studied in Ref. \onlinecite{PhysRevB.92.245410} 
without the local on-site energy term $\epsilon_0$.

In this paper, we study the effect of quantizing magnetic field on Eq. \eqref{Kham:eq}. 
After determining the spectrum we present a novel method, which is based on the operator form of the Green's function of the system,
which is independent from the chosen gauge or representation (i.e. position or momentum). 
To demonstrate the versatility of our method, we calculate the magneto-optical response of the generalized $\alpha-T_3$ lattice, and reproduce
known results along the way for graphene and the pseudospin-1 case with ease.

\section{The operator of the Green's function for the generalized $\alpha-T_3$ model}
\label{landau_levels:sec}

To obtain the magneto-optical conductivity tensor $\text{\boldmath$\sigma$}(\omega)$   
in magnetic field perpendicular to the plane of the dice lattice one needs to calculate the Landau levels (LLs). 
As a standard procedure, replacing the canonical momentum by a gauge-invariant quantity
$\hbar \mathbf{k} \to \text{\boldmath$\Pi$} = \hbar  \mathbf{k}  + |e|  \mathbf{A} $ 
one finds the commutation relation $\left[\Pi_x,\Pi_y\right] = -i \hbar^2/l^2_B$, where $l_B = \sqrt{\frac{\hbar}{e\left| B\right|}}$ 
is the magnetic length scale, and $ \mathbf{A}$ is the vector potential such that 
$ \mathbf{B} = \text{\boldmath$\nabla$} \times \mathbf{A}$. 
By introducing the bosonic creation-annihilation operators 
$\hat{a}= \frac{l_B}{\hbar \sqrt{2}}\, \left(\Pi_x  - i \Pi_y\right)$ 
and $\hat{a}^{\dag} = \frac{l_B}{\hbar \sqrt{2}}\, \left(\Pi_x  + i \Pi_y \right)$ we have $[a,a^{\dag}] = 1$, 
and the Hamiltonian in Eq.~(\ref{Kham:eq}) becomes 
\begin{equation}
\label{HK_B:eq}
H =\left(
\begin{array}{ccc}
0 & \alpha \, \hat{a} & 0 \\ 
\alpha \, \hat{a}^{\dag} & \epsilon_0 & \beta \, \hat{a} \\
0 & \beta \, \hat{a}^{\dag} & 0\\
\end{array}
 \right),
\end{equation}
where $\alpha=\left(3c/\sqrt{2}\right) \, t_1/l_B$ and $\beta=\left(3c/\sqrt{2}\right) \, t_2/l_B$ 
are the rescaled hopping elements $t_1$ and $t_2$, respectively. 

Inspecting the Hamiltonian we assume that the eigenstate is of the form
\begin{align}
\label{Landau_levels_states:eq}
|n,\zeta\rangle &= {(C_{\zeta,1} |n-1 \rangle, C_{\zeta,2}|n \rangle, C_{\zeta,3}|n+1\rangle )}^T, 
\end{align}
where $|n \rangle$ is an eigenstate of the number operator $\hat{N}=\hat{a}^\dag \hat{a}$ with $n=0,1,2,\dots$, 
while the band index is denoted by $\zeta=0, \pm 1$,  
and $C_{\zeta,i}$ with $i=1,2,3$ are coefficients to be determined from the eigenvalue problem of Hamiltonian (\ref{HK_B:eq}). 
The Landau levels $E_n^\zeta$ and the corresponding states are given 
in App.~\ref{eigen:app}.
The Landau levels are different at the $K^\prime$ valley but can be obtained from the above eigenvalues 
by the following replacement $(\alpha, \beta) \rightarrow (-\beta, -\alpha)$. 

Now, we derive the Green's function defined by $G(z)=(z-H)^{-1}$. 
In contrast to the usual way where the Green's function is given in position representation, 
we give the operator form of the Green's function which is independent of any representation.  
We would like to emphasizes that the operator form of the Green's function provides a great simplification  
in the calculation of different physical quantities 
involving the Green's function such as the magneto-optical conductivity. 
Usually, such quantities are expressed in terms of a trace of the product of the Green's function and other operators 
(in this work see Eq.~(\ref{corr_fn:eq}) as an example). 
Now an accepted procedure is to use the position representation of the Green's function.
However, this approach involves complicated analytical calculations. 
Indeed, for example Gusynin and Sharapov recently have used the position representation of 
the proper-time expression for the electron propagator for graphene~\cite{PhysRevB.73.245411,Gusynin_0953-8984-19-2-026222} 
and bilayer graphene~\cite{PhysRevB.86.075414} in homogeneous magnetic field to calculate 
the magneto-optical conductivity.
Using the Schwinger proper-time method~\cite{PhysRevD.42.2881} 
they derive the Fourier transform of the translation invariant part of the Green's function for single and bilayer graphene 
and presented a rather lengthy and complicated derivation to obtain the trace in the expression of the magneto-optical 
conductivity tensor. 
Finally, the evaluation of this trace including integrals of the generalized Laguerre polynomials 
requires further efforts to obtain analytical results. 
As we demonstrate below in contrast to this approach our results, namely the operator form of the Green's function 
gives an elegant way to calculate the trace using only the usual algebra of the creation and annihilation operators.
We easily carried out the whole calculation for graphene using our method and found the same results presented 
in Refs.~\onlinecite{PhysRevB.73.245411,Gusynin_0953-8984-19-2-026222}. 

To show how effective our method is in this work we calculate the magneto-optical conductivity tensor 
for the generalized $\alpha - T_3$ model. 
To this end we need the operator of the Green's function. 
After a lengthy but straightforward analytical calculation we found for the $K$ valley
(for details see App.~\ref{Green:app}):
\begin{widetext}
\begin{subequations}
\label{Green_function:eq}
\begin{align} 
\label{Green_fv_K:eq}
G_{K}(z) &=\left(
\begin{array}{ccc}
\frac{1}{z}\left[I+ \alpha^2 \, (\hat{N}+1)f_{K}(z,\hat{N}+1)\right] & 
\alpha \, \hat{a} f_{K}(z,\hat{N}) & \frac{\alpha \, \beta}{z}\, \hat{a}^2 f_{K}(z,\hat{N}-1) \\[2ex] 
\alpha\, \hat{a}^{\dag}f_{K}(z,\hat{N}+1) & z f_{K}(z,\hat{N}) & \beta \, \hat{a} f_{K}(z,\hat{N}-1) \\[2ex]
\frac{\alpha\, \beta}{z}\,  \hat{a}^{\dag}{}^2 f_{K}(z,\hat{N}+1) & \beta \, \hat{a}^{\dag} f_{K}(z,\hat{N}) & 
\frac{1}{z}\left[I+ \beta^2 \, \hat{N} f_{K}(z,\hat{N}-1)\right]  \\[2ex]
\end{array}
 \right),  \,\,\, \mathrm{where}\\ 
f_{K}(z,\hat{N}) &=\left[z^2-\epsilon_0 z-\alpha^2 \, \hat{N}-\beta^2 \, (\hat{N}+1)\right]^{-1}, 
\end{align}
\end{subequations}
\end{widetext}
while $\hat{N} = \hat{a}^{\dag} \hat{a}$ is the number operator, and $I$ is the identity operator.
The operator of the Green's function for the $K^\prime$ valley can be obtained 
by the transformation (\ref{uniter-U:eq}). 
We should emphasize that $f_{K}(z,\hat{N})$ is an operator 
but can easily be calculated in the Fock representation. 
Note that studying the poles of the Green's function we find the same Landau levels that are given in App.~\ref{eigen:app}.

\section{Magneto-optical conductivity}
\label{Magneto_theor:sec}

Using the Kubo formula~\cite{Mahan_book} the magneto-optical conductivity tensor in the bubble approximation 
can be obtained from the operator of the Green's function given by Eq.~(\ref{Green_function:eq}) 
in the following way
\begin{subequations}
\begin{align} 
\label{sigma_tensor:eq}
\sigma_{\alpha\beta}(\xi) &=
\frac{\Pi_{\alpha\beta}( \xi)-\Pi_{\alpha\beta}(0)}{\xi}, \,\,\, \mathrm{where} \\
\Pi_{\alpha\beta}(i\nu_m) &=\frac{i k_\mathrm{B} T}{2\pi  l_B^2}\, 
\sum\limits_{k=-\infty}^{\infty} \text{Tr}\left(j_\alpha G(i\omega_k+i\nu_m)j_\beta G(i\omega_k)\right). 
\label{corr_fn:eq}
\end{align}%
\end{subequations}%
Here $\Pi_{\alpha\beta}$ is the current-current correlation function ($\alpha, \beta = x,y $), 
$\omega_k=(2k+1) \pi k_\mathrm{B} T$ are the fermionic Matsubara frequencies 
(here $k_\mathrm{B}$ is the Boltzmann constant, $T$ is the temperature and $k$ is an integer) 
and $\nu_m=2 m \pi k_\mathrm{B} T$ are bosonic Matsubara frequencies ($m$ is an integer). 
The trace can be obtained using the eigenstates of the Landau levels given 
in App.~\ref{eigen:app}.
The sum over the fermionic Matsubara frequencies $\omega_k$ in (\ref{corr_fn:eq}) 
can be performed by the usual summation method~\cite{Mahan_book}.
Finally, the current density operator  
$\mathbf{j} = \frac{e}{\hbar}\, \frac{\partial H_K}{\partial \mathbf{k}}$
with Hamiltonian (\ref{Kham:eq}) at the $K$ valley is given by 
\begin{align}
\label{current-op_jx_jy:eq}
j_x &=\frac{e l_B}{\sqrt{2}\hbar}\left(
\begin{array}{ccc}
0 & \alpha  & 0 \\ 
\alpha  & 0 & \beta  \\
0 & \beta  & 0\\
\end{array}
 \right), \,\,\, 
j_y =\frac{i e l_B}{\sqrt{2}\hbar}\left(
\begin{array}{ccc}
0 & -\alpha  & 0 \\ 
\alpha  & 0 & -\beta  \\
0 & \beta  & 0\\
\end{array}
 \right), 
\end{align}
while at the $K^\prime$ valley it is given by the transformation (\ref{uniter-U:eq}).

Then, the frequency dependent magneto-optical conductivity tensor $\text{\boldmath$\sigma$}(\omega)$ 
can be calculated from Eq.~(\ref{sigma_tensor:eq})  
using the usual analytic continuation~\cite{Mahan_book}
$i \nu_m \to \omega + i \eta$ in the current-current correlation function $\Pi_{\alpha\beta}(i\nu)$ 
given by Eq.~(\ref{corr_fn:eq}), where $\eta$ is the inverse life time of the particle.

\section{Results: the magneto-optical conductivity}
\label{Result_magneto:sec}

In this section we present our results for the magneto-optical conductivity. 
The analytical calculation can be carried out in a simple way 
using the algebra of the creation and annihilation operators. 
Our results show explicitly the different contributions to the conductivity 
corresponding to the interband and intraband transitions 
between the flat band and a cone, and between cones in each valley. 
Below the dependence of the conductivity on the frequency, the temperature, the magnetic field and the Fermi energy 
will be discussed.  
Moreover, from these results we shall establish the selection rules for the possible optical excitations 
between Landau levels. 
First, we consider the longitudinal conductivity.

\subsection{The longitudinal conductivity}
\label{Result_sigma_xx_xy:sec}

The total longitudinal conductivity can be written as the sum of terms corresponding 
to intraband and interband transitions. 
After a lengthy but straightforward analytical calculation we find 
\begin{align}
\label{sigmaxx2:eq}
\sigma_{xx}(\omega) &=\sum\limits_{\zeta=\pm}\left(\sigma_{xx,f-c}^{K,\zeta}+\sigma_{xx,c-c,\text{inter}}^{K,\zeta}
+\sigma_{xx,c-c,\text{intra}}^{K,\zeta}\right) \nonumber \\ 
&+(\alpha^2\leftrightarrow\beta^2),
\end{align}
where $\sigma_{xx,f-c}^{K,\zeta}$, $\sigma_{xx,c-c,\text{inter}}^{K,\zeta}$ 
and $\sigma_{xx,c-c,\text{intra}}^{K,\zeta}$ are the contributions to the total longitudinal conductivity 
from the interband transitions between the flat band and a cone, 
the interband transitions between cones, 
and the intraband transitions (within the cones) in the $K$ valley, respectively 
and are given in App.~\ref{sigma_xx_xy:app}.
The contribution to the conductivity from the $K^\prime$ valley is given 
by the second term in (\ref{sigmaxx2:eq}) 
indicated by the replacement $\alpha^2\leftrightarrow\beta^2$. 

\begin{figure}[hbt]
\includegraphics[width=\columnwidth]{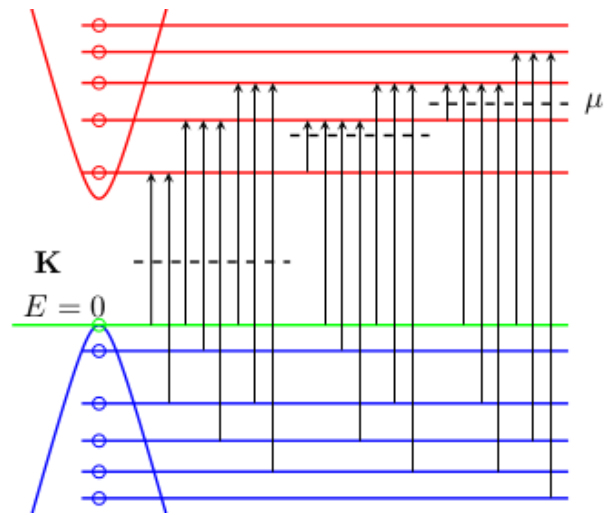}
\caption{(Color online) Allowed transitions for different values of the Fermi energy. 
}
\label{fig: trans}
\end{figure}

\begin{figure}[h]
  \includegraphics[width=\columnwidth]{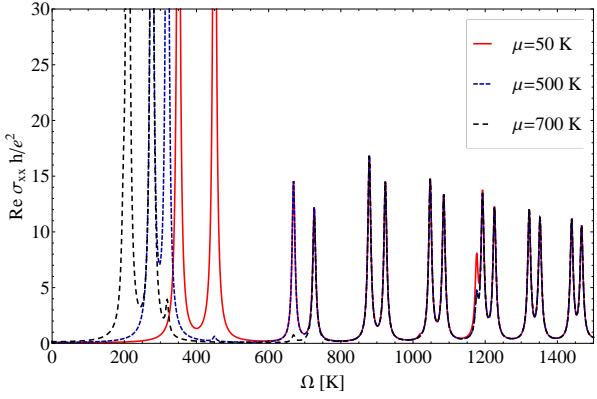}
\caption{(Color online) The real part of the longitudinal conductivity (in units of $e^2/h$) 
as a function of the frequency $\Omega=\hbar\omega/k_{\mathrm{B}}$ (in units of K)
for Fermi energy $\mu=50$ K ((red solid line)) which is in the gap,
$\mu=600$ K (blue short dashed line) which is between 
the Landau levels $n=0$ and $n=1$, and 
$\mu=700$ K (black long dashed line) which lies between 
the Landau levels $n=0$ and $n=1$ in the $K$ valley and 
between the Landau levels $n=1$ and $n=2$ in the $K^\prime$ valley. 
The parameters are $T=10$ K, $\epsilon_0=0$, $\alpha=350$ K, $\beta=450$ K and $\eta=5$ K. }
\label{fig:xxomega}
\end{figure}

To see the allowed transitions between different Landau levels we consider the three contributions to the conductivity 
given by Eq.~(\ref{sigmaxx2:eq}). 
The first term corresponds to the transition from flat band ($\zeta =0$) to cone ($\zeta =1$) and   
at zero temperature the difference of the two Fermi functions becomes nonzero 
if the Landau level indices $n$ of the two energy levels differ exactly by one. 
The magnitude of this contribution is governed by the prefactor. 
For finite temperature in principle other types of transitions are also allowed but much smaller than the ones mentioned above.
The other selection rules can be obtained  from the second and third terms in the expression of the conductivity. 
Analyzing the amplitudes of the different contributions 
it can be shown that the main contribution to the conductivity is the one corresponding to the transition from flat band to the cone band. 
In summary, in Fig.~\ref{fig: trans} we illustrate the allowed transitions for different Fermi energies. 
Figure~\ref{fig:xxomega} shows the conductivities as a function of the frequency $\Omega=\hbar\omega/k_{\mathrm{B}}$ 
for three different chemical potential $\mu$. 
For $\mu = 50$~K the transition $|n=1,\zeta =0\rangle  \to |n=0,\zeta =1\rangle $ gives the two largest peaks in the conductivity corresponding 
to the two valleys. 
While in case of $\mu = 500$~K the transition $|n=0,\zeta =1\rangle  \to |n=1,\zeta =1\rangle $ provides the largest peaks 
in the conductivity. 
Finally, for  $\mu = 700$~K the Landau level indices change as $n=1 \to n=2$ for $K$ valley and $n=0 \to n=1$ for $K^\prime$ valley 
but the quantum number $\zeta=1$ does not change. 

Finally, we discuss the dependence of magneto-optical conductivity on external field. 
It is clear that for  $B\to \infty$ the conductivity should vanish since the distance between the Landau levels tend to infinity.  
The formula for the low field limit is obtained by introducing the variable $\Omega=E_n^{+}-\epsilon_0/2$
and replacing the summation over $n$ into a integral as follows:
\begin{widetext}
\begin{align}
\label{eq: xxlowfield}
&\sigma_{xx}=\frac{2ie^2\xi}{h} \int\limits_{\epsilon_0/2}^{\infty} \mathrm{d}\Omega \left\{\frac{\left(\frac{\epsilon_0}{2}\right)^2+\Omega^2\cos^2(2\phi)}{\Omega^2}\frac{n_\mathrm{F}
\left(\frac{\epsilon_0}{2}-\Omega\right)-n_\mathrm{F}\left(\frac{\epsilon_0}{2}+\Omega\right)}{\xi^2-4\Omega^2}+\right. \nonumber\\
& \qquad \left. {} \sin^2(2\phi)\left[\frac{n_\mathrm{F}(0)-n_\mathrm{F}\left(\frac{\epsilon_0}{2}+\Omega\right)}{\xi^2-\left(\frac{\epsilon_0}{2}+\Omega\right)^2}-\frac{n_\mathrm{F}(0)-n_\mathrm{F}\left(\frac{\epsilon_0}{2}-\Omega\right)}{\xi^2-\left(\frac{\epsilon_0}{2}-\Omega\right)^2}\right]+\frac{\Omega^2-\left(\frac{\epsilon_0}{2}\right)^2}{\Omega\xi^2}\left[\frac{\partial n_\mathrm{F}\left(\frac{\epsilon_0}{2}
-\Omega\right)}{\partial \Omega}
-\frac{\partial n_\mathrm{F}\left(\frac{\epsilon_0}{2}+\Omega\right)}{\partial \Omega}
\right] \right\}, 
\end{align}
\end{widetext}
where $\tan \phi= t_2/t_1$.
In case of $\phi=\pi/4$ (i.e. when $t_1=t_2$) and $\epsilon_0=0$, 
Eq.~(\ref{eq: xxlowfield}) transforms into Eq.~(21) of Ref.~\onlinecite{PhysRevB.84.195422} 
and in case of $\phi=\pi/2$ (graphene) and $\epsilon_0=0$ into Eq.~(13) of Ref.~\onlinecite{Gusynin_0953-8984-19-2-026222}.

\begin{figure}
  \includegraphics[clip,width=\columnwidth]{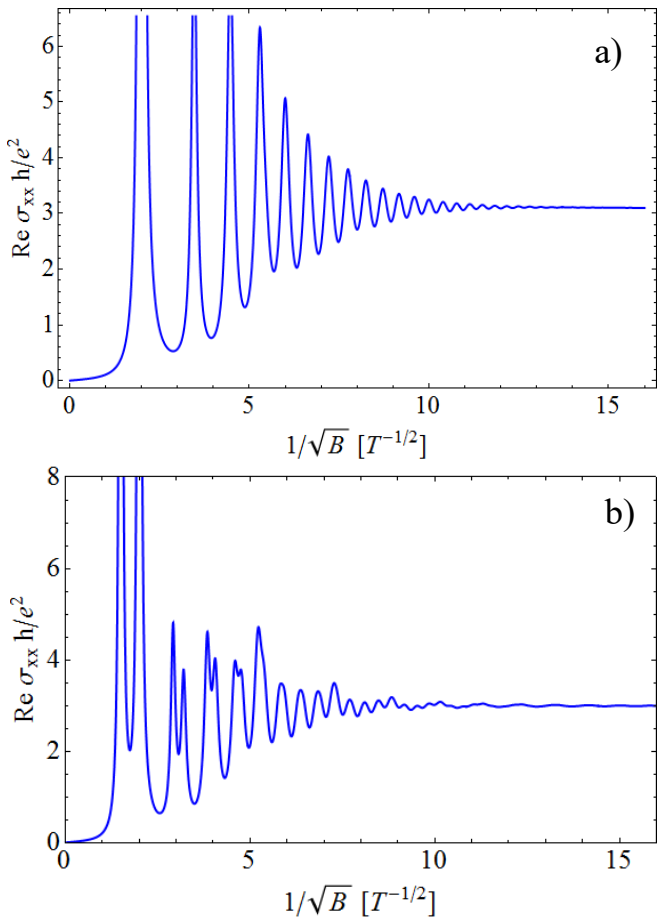}%
\caption{The real part of the longitudinal conductivity (in units of $e^2/h$) 
as a function of the inverse square root of magnetic field $B$ (here $B$ is in units of $\mathrm{T}$) for 
a) $\alpha=\beta=400~\mathrm{K} \cdot \sqrt{B}$,   
b) $\alpha=300~\mathrm{K} \cdot \sqrt{B}$ and $\beta=400~\mathrm{K} \cdot \sqrt{B}$. 
The parameters are $T=10$ K, $\mu=50$ K, $\omega=200$ K, $\eta=5$ K and $\epsilon_0=0$ in both cases. }
\label{fig: xxb}
\end{figure}

As far as intermediate magnetic fields are concerned, 
the height of the peaks and their positions 
can be determined from the results given by Eqs.~(\ref{sigma_xx:eq}).
For simplicity, here we only consider the case $\alpha=\beta$. 
In fact, the pattern for general hopping amplitudes is rather cumbersome as peak energies 
corresponding to different transitions might coincide (approximately) 
and producing a higher peak together (see Fig.~\ref{fig: xxb}), 
even for very small values of the scattering rate.
Let us consider the case when $0 < \mu < E_0^{+}$ so that we do not have to deal with the single intraband transition. 
It is also allowed to neglect cone-to-cone interband peaks according to the arguments above. 
However, we should be careful when we consider the low field limits since 
in this case the main contribution to the peaks in the conductivity results from more than one transitions between the LLs. 
The value of the  real part of the longitudinal conductivity tends to the low magnetic field limit that can be determined 
from the integral in Eq.~(\ref{eq: xxlowfield}). 

One can show that the main characteristics of the oscillation of the longitudinal conductivity as a function of the magnetic field 
is mainly governed by the transitions between the flat band to cone levels.  
For a fixed value of frequency $\omega$ the $mth$ peak (starting from the left hand side in Fig.~\ref{fig: xxb}) occurs 
at $B_m^{-1/2}=\frac{\gamma}{\omega}\sqrt{2m+1}$, where $\gamma = \alpha/ \sqrt{B}$ independent of the magnetic field. 
So the distance between peaks decreases as the difference of the square root of two neighboring odd numbers. 
While for the amplitude of the oscillations one finds
\begin{subequations}
\begin{align}
\label{eq:xxfc1}
& \text{Re}~\sigma_{xx}(\omega = E_m^{+}) \approx \nonumber \\
& \frac{e^2}{\eta h}\left( \frac{\alpha^2+\beta^2+\frac{(m+1)\alpha^4}{E_{m+1}^{+}E_{m+1}^{-}}+\frac{m\beta^4}{E_{m-1}^{+}E_{m-1}^{-}}}{E_m^{+}-E_m^{-}} +(\alpha^2\leftrightarrow\beta^2)\right),
\end{align}
which in case of $\alpha=\beta$ reads as
\begin{equation}\label{eq:xxfc2}
\text{Re}~\sigma_{xx}(\omega=E_m^{+}) \approx \frac{e^2\alpha^2}{\eta h E_m^{+}} \frac{4m^2+4m-5}{4m^2+4m-3}. 
\end{equation}
\end{subequations}
Then from Eq.~(\ref{eq:xxfc2}) it follows that for large enough magnetic field the peaks in the longitudinal conductivity tends to 
\begin{equation}
\text{Re}~\sigma_{xx}(\omega=E_m^{+}) \approx  \frac{e^2}{\eta h}\, \frac{\gamma^2 B_m}{\omega},
\end{equation}
which is proportional to the position of the peaks $B_m$.

\subsection{The transversal conductivity}
\label{Result_sigma_xy:sec}

Similarly to the case of longitudinal conductivity the transversal conductivity 
(off-diagonal component of the conductivity tensor $\text{\boldmath$\sigma$}$)
can also be written as the sum of terms corresponding to intraband and interband transitions. 

After a lengthy but straightforward analytical calculation we find 
\begin{align}
\label{sigmaxy2:eq}
\sigma_{xy}(\omega) &=\sum\limits_{\zeta=\pm}
\left(\sigma_{xy,f-c}^{K,\zeta}+\sigma_{xy,c-c,\text{inter}}^{K,\zeta}
+\sigma_{xy,c-c,\text{intra}}^{K,\zeta}\right) \nonumber \\
&+(\alpha^2\leftrightarrow\beta^2), 
\end{align}
where $\sigma_{xy,f-c}^{K,\zeta}$, $\sigma_{xy,c-c,\text{inter}}^{K,\zeta}$ 
and $\sigma_{xy,c-c,\text{intra}}^{K,\zeta}$ are the contributions to the total transversal conductivity 
from the interband transitions between the flat band and a cone, 
the interband transitions between cones, 
and the intraband transitions (within the cones) in the $K$ valley, respectively 
and are given in App.~\ref{sigma_xx_xy:app}.
The contribution to conductivity from the $K^\prime$ valley is given by the second term 
in (\ref{sigmaxy2:eq}) indicated by the replacement $\alpha^2\leftrightarrow\beta^2$. 

\begin{figure}[htb]
  \includegraphics[clip,width=\columnwidth]{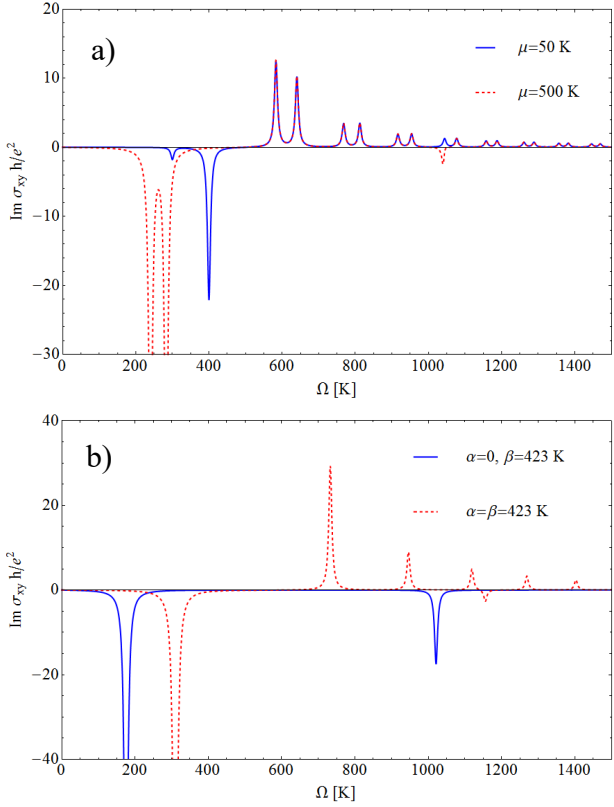}%
\caption{(Color online) The imaginary part of the transversal conductivity (in units of $e^2/h$) 
as a function of the frequency $\Omega=\hbar\omega/k_{\mathrm{B}}$ (in units of K)
a) for Fermi energy $\mu=50$~K lying between the flat band and the first LL (blue solid line) 
and $\mu=500$~K which is between the first and second LL (red dashed line), 
b) for $\mu=50$ K and for two sets of hopping parameters: $\epsilon_0=\alpha=0$, $\beta=423$ K 
corresponding to graphene with $B=1$ T magnetic field (blue solid line), and 
$\epsilon_0=0$, $\alpha=\beta=423$ K related to the Dirac-Weyl model for $s=1$ (red dashed line). 
The parameters are  $T=10$ K and $\eta=5$ K in both cases.  
}
\label{fig:xyomega}
\end{figure}

Figure \ref{fig:xyomega} shows the Hall conductivity (the imaginary part of the off-diagonal component of the conductivity tensor) as a function the frequency 
for different chemical potential (in panel a) and for different hopping amplitudes (in panel b). 
When $0< \mu < E_0^{+}$ in a valley (blue solid line) then there is no intraband transition 
so peaks in the conductivity result only from flat band to cone and cone to cone transitions. 
In this case there is a negative peak (around $\Omega \approx 400$ K in the figure) corresponding to the transition 
$|n=1,\zeta =0\rangle\rightarrow |n=0,\zeta= 1\rangle$. 
All the other peaks corresponding to other flat band to cone transitions are positive. 
Small negative peaks (around $\Omega \approx 300$ K in the figure) due to interband cone to cone transitions are also present. 
However, if  $E_0^{+} < \mu <  E_1^{+}$ (red dashed line) then the aforementioned negative peak from flat band 
to cone transition disappears, while another negative peak appears due to an intraband transition. 

The heights of positive peaks fall rapidly in both cases in terms of the frequency according to 
\begin{subequations}
\begin{align}
\label{eq:xyfc1}
& \text{Im}~\sigma_{xy}(\omega=E_m^{+})\approx \nonumber \\
& \frac{e^2}{\eta h}\left( \frac{\beta^2-\alpha^2+\frac{m\beta^4}{E_{m-1}^{+}E_{m-1}^{-}}-\frac{(m+1)\alpha^4}{E_{m+1}^{+}E_{m+1}^{-}}}{E_m^{+}-E_m^{-}}+(\alpha^2\leftrightarrow\beta^2)\right),
\end{align}
which in case of $\alpha=\beta$ reads as
\begin{equation}\label{eq:xyfc2}
\text{Im}~\sigma_{xy}(\omega=E_m^{+})\approx 
\frac{e^2\alpha^2}{\eta h E_m^{+}} \frac{2m+1}{4m^2+4m-3}.
\end{equation}
\end{subequations}

\begin{figure}[h]
  \centering
  \includegraphics[width=\linewidth]{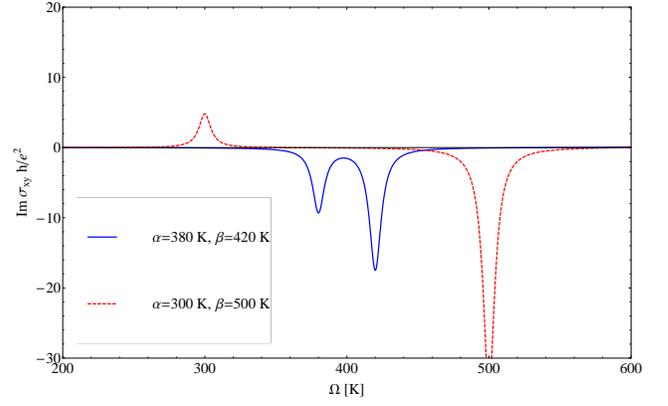}
\caption{(Color online) The imaginary part of the transversal conductivity (in units of $e^2/h$) 
as a function of the frequency $\Omega=\hbar\omega/k_{\mathrm{B}}$ (in units of K), 
for hopping amplitudes $\epsilon_0=0$, $\alpha=380$ K, $\beta=420$ K (blue solid line), and 
for $\epsilon_0=0$, $\alpha=300$ K, $\beta=500$ K (red dashed line).
The parameters are $T=10$ K and $\eta=5$ K. }
\label{fig:xyomega4}
\end{figure}

It is also worth noting that when $\alpha\neq\beta$ 
then the first negative flat band to cone peak splits into two peaks 
(corresponding to $K$ and $K'$ valleys, respectively) as shown in Fig.~\ref{fig:xyomega}a around $\Omega \approx 250$ K. 
If the difference between the two hopping amplitudes are large enough then it might occur 
that one of these two peaks becomes positive as can be seen in Fig.~\ref{fig:xyomega4} exactly at $\Omega = 300$ K.

Finally, we consider the transversal conductivity  in the DC limit ($\omega = 0$) and at zero temperature. 
In this case we obtain the usual Hall conductivity. 
From Eqs.~(\ref{sigma_xy:eq}) we can find the contribution from the $K$ and $K^\prime$ valleys as   
\begin{subequations}
\label{sxy_K-and-K':eq}
\begin{align}
\sigma_{xy}^{K} &=\frac{e^2}{h}\left(\frac{\alpha^2-\beta^2}{\alpha^2+\beta^2}F(0)-2\sum\limits_{n=0}^{\infty}F_n^{K}\right), \\
\sigma_{xy}^{K'} &=\frac{e^2}{h}\left(\frac{\beta^2-\alpha^2}{\alpha^2+\beta^2}F(0)-2\sum\limits_{n=0}^{\infty}F_n^{K^\prime}\right), 
\end{align}
\end{subequations}
where $F_n^K=n_\mathrm{F}(E_n^{+})+n_\mathrm{F}(E_n^{-})$ and $E_n^{\pm}$ are the energy levels for the $K$ valley, and 
$F_n^{K^\prime}$ is the same as  $F_n^K$ with energy levels for the $K^\prime$ valley. 
Here the spin degeneracy is taken into account. 
Thus the total contributions from $K$ and $K^\prime$ valleys can be rewritten as 
\begin{equation}
\sigma_{xy}(\mu,B,T)= \sigma_{xy}^{K}+\sigma_{xy}^{K'} =-\frac{2e^2}{h}\sum\limits_{n=0}^{\infty}\left(F_n^{K}+F_n^{K^\prime}\right).
\end{equation}

\begin{figure}[htb]
  \includegraphics[clip,width=\columnwidth]{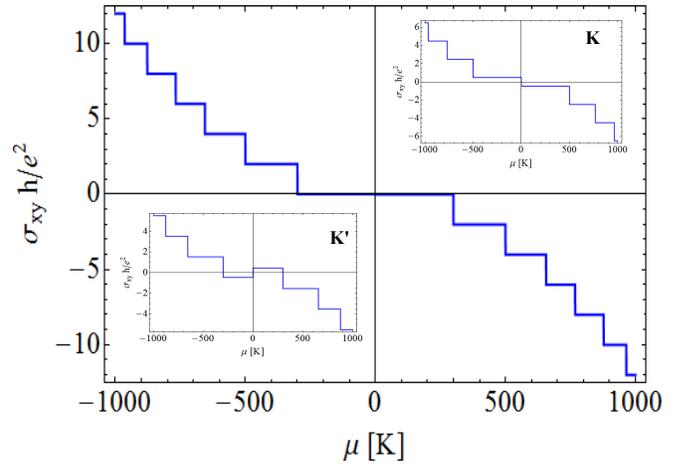}%
\caption{The transversal conductivity (in units of $e^2/h$) in DC limit ($\omega =0$) 
as a function of the Fermi energy. 
Insets (upper right and lower left) show the contributions from the $K$ and $K^\prime$ valleys, respectively. 
The parameters: $T=0.01$ K, $\eta= 5$ K, $\epsilon_0=0$, $\alpha=300$ K, $\beta=500$ K.
}
\label{fig: hall}
\end{figure}

\begin{figure}[htb]
  \includegraphics[clip,width=\columnwidth]{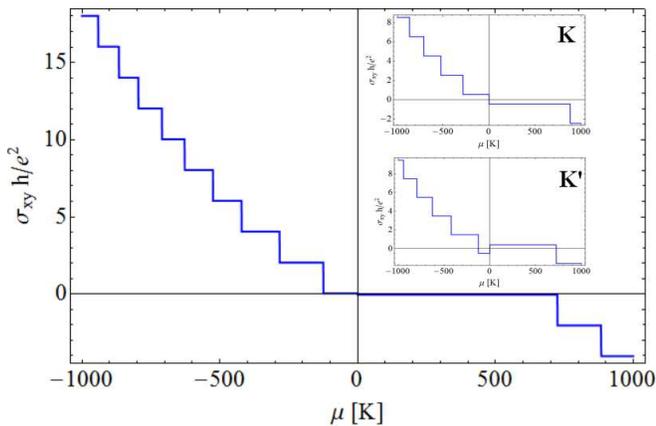}%
\caption{The transversal conductivity (in units of $e^2/h$) in DC limit ($\omega =0$)
as a function of the Fermi energy.
Insets (upper right and lower left) show the contributions from the $K$ and $K^\prime$ valleys, respectively.
The parameters: $T=0.01$ K, $\eta= 5$ K, $\epsilon_0=600$ K, $\alpha=300$ K, $\beta=500$ K.
}
\label{fig: hall3}
\end{figure}

The Hall conductivity as a function of the Fermi energy at zero temperature and in DC limit is plotted in Figs.~\ref{fig: hall} and ~\ref{fig: hall3}.
The insets in this figure show the individual contributions from the two valleys to the conductivity. 
It can be shown that the conductivity is zero when $\mu$ is in the narrower energy interval  $\{E_0^- ,E_0^+ \}$ corresponding to the $K$ and $K^\prime$ valleys. 
For parameters used in the figure this is $|\mu |< \alpha$. 
Moreover, the conductivity has a change $2 e^2/h$  at all the other Landau levels. 
For $\epsilon_0\neq 0$, the Hall conductivity looses its symmetry with respect to $\mu=0$.
Our results for $\epsilon_0=0$ agree with those on Ref. \onlinecite{biswas}.

We now show that this Hall conductivity can be related to the Berry phase. 
Indeed, when the temperature is zero and the Fermi energy lies between the flat band and first LL then 
the sums in Eq.~(\ref{sxy_K-and-K':eq}) becomes zero and $F(0) =2$. 
Then the conductivity becomes
\begin{align}
 \sigma_{xy}^{K,K'} &= \pm 2\, \frac{e^2}{h}\, \frac{\alpha^2-\beta^2}{\alpha^2+\beta^2} = \pm 2\, \frac{e^2}{h}\cos(2\phi),
\end{align}
where $\phi$ is given by $\tan \phi = t_2/t_1$ and the spin degeneracy is included.  
This results is in agreement with the Berry phases obtained in Refs.~\onlinecite{PhysRevLett.112.026402,PhysRevB.92.245410}. 
It is interesting to note that the valley resolved Hall response is not only \emph{fractional} but can also be rational without any electron-electron interactions, albeit the 
sum of the two valleys, the total Hall response is always integer quantized.

\section{Conclusions}
\label{conlusion:sec}

In this work the magneto-optical conductivity in the generalized $\alpha - T_3$ model is calculated. 
In this generalized form we assumed that the on-site energy $\epsilon_0$ of the sixfold connected site 
can be non-zero.
Using the Kubo formula expressed with Green's function, the magneto-optical conductivity tensor 
is calculated as functions of frequency, external field, temperature and Fermi energy. 
To this end we introduce a new analytical procedure to determine the Green's function 
in an operator form independent of any representation.
When the Green's function is given in position representation the evaluation of the trace 
in the Kubo formula is a quite cumbersome analytical calculation. 
The advantage of our approach is that the Kubo formula can be calculated in simple way 
using only the algebra of the creation and annihilation operators.  
To demonstrate the theoretical method mentioned above, the calculations are also carried out 
for graphene and it is shown that the results obtained from our new method are in agreement 
with those known in the literature. 

From our general result for the transversal conductivity we derived an analytic expression 
for the Hall-conductivity in DC limit. 
We show that the Hall conductivity at zero temperature agrees with 
that obtained from the Berry phase calculated in earlier works. 
Moreover, the Hall conductivity is integer quantized and 
the steps of quanta depend continuously on the hopping parameters between adjacent layers. 

We believe that our predictions for the magneto-optical conductivity can be tested experimentally with cold atoms in an optical lattice. 
Furthermore, our new algorithm is an efficient and universal approach and thus easily applicable to other systems.

\acknowledgments 

We would like to thank  A. P\'alyi  for helpful discussions.
This work is supported by The National Research, Development and Innovation Office under the contracts Nos.~ 101244, 105149 and 108676.

\appendix

\section{Eigenvalues and eigenstates of the systems}
\label{eigen:app}

In this section we present the eigenvalues and the eigenstates of the Hamiltonian (\ref{HK_B:eq}) (around the $K$ valley). 
The Schr\"odinger equation reads as 
\begin{align}
\label{Sch_Landau:eq}
 H |n,\zeta\rangle &= E_n^\zeta |n,\zeta\rangle,  
\end{align}
where $E_n^\zeta$ and $|n,\zeta\rangle$ is the energy eigenvalue and the corresponding eigenstate, respectively, 
and $n=0,1,2, \dots$ and $\zeta=-1,0,+1$ denote the Fock number and the band index, respectively. 
To solve this equation, we look for a solution of the form given by Eq.~(\ref{Landau_levels_states:eq}). 
The results are summarized in Table~\ref{Landau_levels_states:table}. 
\begin{widetext}

\begin{table}[htb]
\centering
\begin{tabular}{ccc}
\hline
\hline
$n$, $\zeta$  & $E_n^\zeta$ & $|n,\zeta\rangle$   \\ 
\hline
 & & \\
$n>0$, $\zeta=\pm 1$ & $E_n^{\pm 1} = \frac{\epsilon_0}{2}+\zeta\sqrt{\left(\frac{\epsilon_0}{2}\right)^2+\alpha^2 n+\beta^2 (n+1)}$  
& ${\gamma_n^{\pm 1} \left(\alpha\sqrt{n} |n-1\rangle, E_n^{\pm 1} |n\rangle ,  \beta\sqrt{n+1} |n+1\rangle \right)}^T$  \\
& & \\
$n>0$, $\zeta=0$ & $E_n^{0} = 0$ & $ {\gamma_n^{0} \left(-\beta\sqrt{n+1} |n-1\rangle, 0 ,  \alpha\sqrt{n} |n+1\rangle \right)}^T$ \\
& & \\
$n=0$, $\zeta=\pm 1$ & $E_0^{\pm 1} = \frac{\epsilon_0}{2}+\zeta\sqrt{\left(\frac{\epsilon_0}{2}\right)^2+\beta^2 }$ 
& ${\gamma_0^{\pm 1} \left(0, E_0^{\pm 1} |0\rangle,  \beta |1\rangle \right)}^T$ \\
& & \\
$n=0$, $\zeta=0$ & $E_0^{0} =0$ &  ${\left(0,0 ,  |0\rangle \right)}^T$\\
\hline
\hline
\end{tabular}
\caption{Landau levels and eigenstates for valley $K$. 
Each levels are labeled by the Fock number $n$ and a band index $\zeta$.
The normalization factors are 
$\gamma_n^{\zeta}=\left(\alpha^2 n+\beta^2 (n+1)+(E_n^{\zeta})^2\right)^{-\frac{1}{2}}$ for $\zeta=\pm$ 
and $\gamma_n^{0}=\left(\alpha^2 n+\beta^2 (n+1)\right)^{-\frac{1}{2}}$ for $\zeta=0$.}
\label{Landau_levels_states:table}
\end{table}

\end{widetext}

\section{Calculation of the Green's function}
\label{Green:app}

To obtain the operator of the Green's function for the Hamiltonian $H$ given by Eq.~(\ref{HK_B:eq}) 
we partitioned the operator $z-H$ as 
\begin{align}
\label{z-H:eq}
z-H &=\left(
\begin{array}{cc|c}
z & -\alpha \, \hat{a} & 0 \\ 
-\alpha \, \hat{a}^{\dag} & z-\epsilon_0 & -\beta \, \hat{a} \\
\hline
0 & -\beta \, \hat{a}^{\dag} & z\\
\end{array}
 \right) 
\equiv 
\left(
\begin{array}{cc}
A & B \\ 
C & D\\
\end{array}
 \right).
\end{align}
Then we apply the general formula for the inverse of a 2 by 2 partitioned matrix
\begin{align}
\label{inverse_formula:eq}
{\left(
\begin{array}{cc}
A & B \\ 
C & D\\
\end{array}
 \right)}^{-1}
 &= \left(
\begin{array}{cc}
A^{-1}+A^{-1}B S^{-1} CA^{-1} & -A^{-1}B S^{-1} \\ 
-S^{-1} CA^{-1} & S^{-1}
\end{array}
 \right),
\end{align}
where $S = D-CA^{-1}B$ and the operators $A$ and $S$ can be inverted.
This is often called in the literature the Banachiewicz inversion 
formula~\cite{Fuzhen_Zhang:book,companion:book}.
 
The inverse of operator $A$ defined in (\ref{z-H:eq}) can also be calculated from formula (\ref{inverse_formula:eq}) 
and after a simple algebra we find
\begin{align}
\label{A_inverse:eq}
 A^{-1} &=\left(
\begin{array}{cc}
(z-\epsilon_0) \, p(z,\hat{N}+1) & \alpha \, \hat{a} \, p(z,\hat{N}) \\ 
\alpha\,\hat{a}^{\dag} \, p(z,\hat{N}+1) & z \, p(z,\hat{N}) \\
\end{array}
 \right), 
\end{align}
where $p(z, \hat{N})=\left({z^2-\epsilon_0 z-\alpha^2 \hat{N}}\right)^{-1}$ 
and $\hat{N} = \hat{a}^{\dag} \hat{a}$ is the number operator.
Now using (\ref{A_inverse:eq}) and the general formula (\ref{inverse_formula:eq}) 
the matrix elements of the inverse of matrix in (\ref{z-H:eq}) can be calculated analytically and we find
\begin{subequations}
\label{inverse_terms:eq}
\begin{eqnarray}
&& S^{-1} = \frac{1}{z}\left(I+\beta^2 N f(z,\hat{N}-1)\right),  \\[2ex]
&& -A^{-1}B S^{-1} = \left(
\begin{array}{c}
 \frac{\alpha\beta}{z}\, \hat{a}^2 f(z,\hat{N}-1)\\ 
 \beta \, \hat{a} f(z,\hat{N}-1) \\
\end{array}
 \right),  \\[2ex]
&& \! \! \! -S^{-1} CA^{-1} = \! \! \left(
\begin{array}{cc}
\frac{\alpha\beta}{z}\, \hat{a}^{\dag}{}^2 f(z,\hat{N}+1), & \beta \, \hat{a}^{\dag} f(z,\hat{N})
\end{array}
\! \!  \right), \\[2ex] 
 \lefteqn{A^{-1}+A^{-1}B S^{-1} CA^{-1} = }   \nonumber \\ 
 && \left(
\begin{array}{cc}
\frac{1}{z}\left[I+ \alpha^2\, (\hat{N}+1)f(z,\hat{N}+1)\right] & \alpha\, \hat{a} f(z,\hat{N})  \\ 
\alpha\, \hat{a}^{\dag}f(z,\hat{N}+1) & z f(z,\hat{N})  \\
\end{array}
 \right),
\end{eqnarray}
\end{subequations}
where $f(z,\hat{N})=\left[z^2-\epsilon_0 z-\alpha^2 \, \hat{N}-\beta^2 \, (\hat{N}+1)\right]^{-1}$.

Here we have made use of the following identities: 
\begin{align} 
\hat{a}\, f(z,\hat{N}) &= f(z,\hat{N}+1)\, \hat{a}^{\dag} \\
\hat{a}^{\dag}\,  f(z,\hat{N}) &= f(z,\hat{N}-1)\, \hat{a}.
\end{align}
Finally, substituting the terms given by Eqs.~(\ref{inverse_terms:eq}) 
into Eq.~(\ref{inverse_formula:eq}) we obtain 
the operator of the Green's function $G(z) = {\left(z-H \right)}^{-1}$ as given by Eq.~(\ref{Green_function:eq}).
For the case of $K^\prime$ valley the Green's function can be obtained 
by the transformation (\ref{uniter-U:eq}).

\section{Expressions for the longitudinal and transversal conductivities}
\label{sigma_xx_xy:app}

Using the operator form of the Green's function given by Eq.~(\ref{Green_function:eq}) 
and the current operators (\ref{current-op_jx_jy:eq}), 
and performing the Matsubara summation in (\ref{corr_fn:eq})  the magneto-optical conductivity can be calculated analytically. 
Then the longitudinal conductivity is given by Eq.~(\ref{sigmaxx2:eq}) in which the different terms reads
\begin{widetext}
\begin{subequations}
\label{sigma_xx:eq}
\begin{align}
\sigma_{xx,f-c}^{K,\zeta}(\omega) &=\frac{ie^2}{h}\sum\limits_{n=0}^{\infty} 
\frac{\alpha^2+\beta^2+\frac{(n+1)\alpha^4}{E_{n+1}^{+}E_{n+1}^{-}}+\frac{n\beta^4}{E_{n-1}^{+}E_{n-1}^{-}}}{E_n^\zeta-E_n^{-\zeta}}\left(\frac{1}{\xi-E_n^{\zeta}}+\frac{1}{\xi+E_n^{\zeta}} \right)
\left[n_\mathrm{F}(0)-n_\mathrm{F}(E_n^\zeta) \right],  \\
\sigma_{xx,c-c,\text{inter}}^{K,\zeta}(\omega) &=\frac{ie^2}{h}\sum\limits_{n=0}^{\infty} 
\frac{ (n+1)\left(\alpha^2 E_n^{\zeta}+\beta^2 E_{n+1}^{-\zeta}\right)^2}{\left( E_n^{\zeta}-E_n^{-\zeta} \right)
\left( E_{n+1}^{\zeta}-E_{n+1}^{-\zeta} \right) \left(- E_n^{\zeta}E_{n+1}^{-\zeta} \right)  
\left(E_{n}^{\zeta}-E_{n+1}^{-\zeta} \right) }\times  \\ \nonumber
&\left( \frac{1}{\xi+E_n^{\zeta}-E_{n+1}^{-\zeta}}+\frac{1}{\xi-E_n^{\zeta}+E_{n+1}^{-\zeta}}\right)
\left[n_\mathrm{F}(E_{n+1}^{-\zeta})-n_\mathrm{F}(E_n^\zeta) \right], \\ 
\sigma_{xx,c-c,\text{intra}}^{K,\zeta}(\omega) &=\frac{ie^2}{h}\sum\limits_{n=0}^{\infty} \frac{ (n+1)\left(\alpha^2 E_n^{\zeta}+\beta^2 E_{n+1}^{\zeta}\right)^2}{\left( E_n^{\zeta}-E_n^{-\zeta} \right)
\left( E_{n+1}^{\zeta}-E_{n+1}^{-\zeta} \right) \left( E_n^{\zeta}E_{n+1}^{\zeta} \right) 
\left(E_{n+1}^{\zeta}-E_{n}^{\zeta} \right) }\times  \\ \nonumber
&\left( \frac{1}{\xi+E_n^{\zeta}-E_{n+1}^{\zeta}}+\frac{1}{\xi-E_n^{\zeta}+E_{n+1}^{\zeta}}\right)
\left[n_\mathrm{F}(E_{n}^{\zeta})-n_\mathrm{F}(E_{n+1}^\zeta) \right],  
\end{align}
\end{subequations}
Similar calculations leads to the Hall conductivity given by Eq.~(\ref{sigmaxy2:eq}) in which the different terms are  
\begin{subequations}
\label{sigma_xy:eq}
\begin{align}
\sigma_{xy,f-c}^{K,\zeta}(\omega) &=\frac{e^2}{h}\sum\limits_{n=0}^{\infty} \frac{\beta^2-\alpha^2+\frac{n\beta^4}{E_{n-1}^{+}E_{n-1}^{-}}-\frac{(n+1)\alpha^4}{E_{n+1}^{+}E_{n+1}^{-}}}{E_n^\zeta-E_n^{-\zeta}}\left(\frac{1}{\xi-E_n^{\zeta}}-\frac{1}{\xi+E_n^{\zeta}} \right)\left[n_\mathrm{F}(0)-n_\mathrm{F}(E_n^\zeta) \right], \\
\sigma_{xy,c-c,\text{inter}}^{K,\zeta}(\omega) &=\frac{e^2}{h}\sum\limits_{n=0}^{\infty} \frac{ (n+1)\left(\alpha^2 E_n^{\zeta}+\beta^2 E_{n+1}^{-\zeta}\right)^2}{\left( E_n^{\zeta}-E_n^{-\zeta} \right)\left( E_{n+1}^{\zeta}-E_{n+1}^{-\zeta} \right) \left(- E_n^{\zeta}E_{n+1}^{-\zeta} \right)  
\left(E_{n}^{\zeta}-E_{n+1}^{-\zeta} \right) }\times  \\ \nonumber
&\left( \frac{1}{\xi+E_n^{\zeta}-E_{n+1}^{-\zeta}}-\frac{1}{\xi-E_n^{\zeta}+E_{n+1}^{-\zeta}}\right)\left[n_\mathrm{F}(E_{n+1}^{-\zeta})-n_\mathrm{F}(E_n^\zeta) \right], \\
\sigma_{xy,c-c,\text{intra}}^{K,\zeta}(\omega) &=\frac{e^2}{h}\sum\limits_{n=0}^{\infty} \frac{ (n+1)\left(\alpha^2 E_n^{\zeta}+\beta^2 E_{n+1}^{\zeta}\right)^2}{\left( E_n^{\zeta}-E_n^{-\zeta} \right)
\left( E_{n+1}^{\zeta}-E_{n+1}^{-\zeta} \right) \left( E_n^{\zeta}E_{n+1}^{\zeta} \right)  \left(E_{n+1}^{\zeta}-E_{n}^{\zeta} \right) }\times  \\ \nonumber
&\left( \frac{1}{\xi+E_n^{\zeta}-E_{n+1}^{\zeta}}-\frac{1}{\xi-E_n^{\zeta}+E_{n+1}^{\zeta}}\right)\left[n_\mathrm{F}(E_{n}^{\zeta})-n_\mathrm{F}(E_{n+1}^\zeta) \right],
\end{align}
\end{subequations}
\end{widetext}
where $n_\mathrm{F}(E) = 1/(e^{\left(E-\mu \right)/(k_{B}T)}+1)$ is the Fermi distribution function, 
$\mu$ is the Fermi energy and $\xi = \hbar \omega + i\eta$.

\bibliographystyle{apsrev4-1}


%

\end{document}